\documentclass{elsarticle}
\usepackage{graphicx}
\usepackage{bm}
\usepackage{amsfonts,amsmath}
\usepackage{xcolor}
\usepackage[colorlinks=true]{hyperref}

\newcommand{\mmm}{\mathbf m}
\newcommand{\rrr}{\mathbf r}
\newcommand{\RRR}{\mathbf R}

\title{Gyroscopic tensor of a magnetic soliton}
\author{Rocio Gonzalez and Oleg Tchernyshyov}
\ead{olegt@jhu.edu}
\address{William H. Miller III Department of Physics and Astronomy and Institute for Quantum Matter, Johns Hopkins University, Baltimore, MD 21218, USA}
\date{\today}

\begin{document}

\begin{abstract}
The gyroscopic force acting on a ferromagnetic soliton reflects the tendency of spins to precess. Its components increase linearly with (generalized) velocities of the soliton. The proportionality coefficients form the gyroscopic tensor, a generalization of Thiele's gyrovector. Originally introduced as an auxiliary quantity, the gyroscopic tensor turns out to be a mathematical object of fundamental importance with a long history going back to Lagrange. We review the applications of the gyroscopic tensor and its historical roots. 
\end{abstract}

\maketitle

\section{Introduction}

This paper focuses on the gyroscopic tensor, a mathematical object used in the theory of magnetic solitons. Magnetic moments in a ferromagnet come from spins, and spins precess like gyroscopes. The precession of spins in a magnet gives rise to a gyroscopic force acting on a moving soliton. Components of this force are linearly proportional to (generalized) velocities of the soliton. The proportionality coefficients form the gyroscopic tensor. In magnetism, this concept was first introduced in 1973 by Thiele \cite{Thiele:1973a} for a rigidly moving soliton and extended by Clarke et al. \cite{Clarke:2008} to general soliton dynamics. Yet the concept turns out to be much older, going back more than two hundred years to Lagrange \cite{Lagrange:1877}. 

Furthermore, the gyroscopic tensor is more than just a proportionality constant between a velocity and a force. It provides a straightforward way to compute conserved momenta of a soliton, an otherwise complex and laborious process. It also underlies the quasiclassical count of physical states of a soliton and canonical quantization of its dynamics. 

We review the concept and applications of the gyroscopic tensor, which originated as the gyrovector in Thiele's 1973 work on rigidly moving solitons and was later extended to general soliton dynamics. We use simple models of magnetic solitons in 1 and 2 spatial dimensions to illustrate the general theory. We show that the Lagrange bracket, introduced in 1808 in the context of celestial mechanics, is a direct analog of the gyroscopic tensor. We also give a new general formula for counting the physical states of a ferromagnetic soliton, where the gyroscopic tensor plays a prominent role.  

\section{Translational motion}
\label{sec:translational-motion}

At temperatures well below the Curie point and on length scales long compared to the lattice spacing, spins in a ferromagnet can be described by a field theory known as micromagnetics \cite{Brown:1978}. The spin in a mesoscopic volume $dV$ is $\mathcal S \mmm \, dV$, where $\mmm(\mathbf r,t)$ is a vector field of unit length and $\mathcal S \mmm$ is the spin density. The length constraint can be resolved by expressing $\mmm$ in spherical angles,
\begin{equation}
\mmm = 
(\sin{\theta} \cos{\phi}, \,
\sin{\theta} \sin{\phi}, \,
\cos{\theta}).
\label{eq:m-theta-phi}
\end{equation}

The time evolution of the field $\mmm(\mathbf r,t)$ is given by the Landau-Lifshitz equation \cite{Landau:1935}, which equates the rate of change of the local spin density $\mathcal S \mmm$ to the density of torque from conservative forces:
\begin{equation}
\mathcal S \partial_t \mmm = 
- \mmm \times \frac{\delta U[\mmm(\rrr)]}{\delta \mmm(\rrr)}.
\label{eq:LLG}
\end{equation}
Here $U[\mmm(\rrr)]$ is an energy functional; its functional derivative $\delta U[\mmm(\rrr)]/\delta \mmm(\rrr)$ is known as an effective (magnetic) field. We do not consider the effects of dissipation in this paper. 

In a typical ferromagnet, the energy functional is dominated by the Heisenberg exchange energy 
\begin{equation}
U = \int dV \, 
\frac{A}{2} \left[
(\partial_x \mmm)^2 +
(\partial_y \mmm)^2 +
(\partial_z \mmm)^2
\right]
+ \ldots
\end{equation}
The omitted terms include the effects of weaker anisotropic spin interactions resulting from spin-orbit coupling. The exchange energy, quadratic in the gradients of the spin field $\mmm$, makes the Landau-Lifshitz equation a second-order partial differential equation, $\mathcal S \partial_t \mmm = 
A \mmm \times \nabla^2 \mmm + \ldots$, and a nonlinear one on account of the constraint $|\mmm(\rrr,t)| = 1$. Exact solutions describing dynamic solitons are exceedingly rare, so one must resort to approximations in order to make progress. 

A crucial step in this direction was made by Thiele \cite{Thiele:1973a}, who derived equations of motion for a rigidly moving soliton. Suppose the Landau-Lifshitz equation (\ref{eq:LLG}) has a static soliton solution $\mmm_0(\rrr)$ for some energy functional $U_0[\mmm(\rrr)]$. Suppose further that the energy functional respects translational invariance, $U_0[\mmm_0(\rrr-\RRR)] = U_0[\mmm_0(\rrr)]$, so that a translated soliton $\mmm_0(\rrr-\RRR)$ is also a static solution of Eq.~(\ref{eq:LLG}). Let us now add a weak perturbation expressed by a small additional energy $U_1[\mmm(\rrr)]$ that breaks the translational symmetry and induces slow motion of the soliton. Assuming that the soliton is sufficiently rigid, we may guess that a weak perturbation won't change the shape of the soliton $\mmm_0(\rrr)$ and that its motion will be satisfactorily captured by a time evolution of the translation vector $\RRR(t)$: 
\begin{equation}
\mmm(\rrr,t) \approx \mmm_0(\rrr-\RRR(t)). 
\end{equation}
Thiele derived an equation for the soliton velocity $\dot{\RRR}$,
\begin{equation}
\dot{\RRR} \times \mathbf G - \partial U_1/\partial \RRR = 0,
\label{eq:Thiele}
\end{equation}
where $U_1 = U_1[\mmm_0(\rrr-\RRR)]$ has become a function of $\RRR$. Thiele's equation (\ref{eq:Thiele}) has a new quantity, the gyrovector $\mathbf G = (G^x, G^y, G^z)$ with Cartesian components
\begin{equation}
G^i = 
- \frac{1}{2} \sum_{j,k} \epsilon^{ijk}   
\int dV \, \mathcal S 
\mmm \cdot 
(\partial_j \mmm \times \partial_k \mmm).
\label{eq:gyrovector}
\end{equation}
Here $i = x, y, z$ and $\epsilon^{ijk}$ is the fully antisymmetric Levi-Civita symbol. 

Eq.~(\ref{eq:Thiele}) expresses the balance of forces acting on the soliton: a conservative force originating from the potential energy landscape $U_1(\RRR)$ and the gyroscopic force $\dot{\RRR} \times \mathbf G$ resembling the Lorentz force acting on a charged particle in a magnetic field.   

\section{General soliton motion}

Thiele's approach can be generalized beyond rigid translations. The dynamics of a soliton $\mmm(\rrr, \bm \xi)$ parametrized by a set of collective coordinates 
\begin{equation}
\bm \xi \equiv \{\xi^1, \xi^2, \ldots, \xi^N\}
\label{eq:xi-collective-coordinates}
\end{equation}
can be described through the time evolution $\bm \xi(t)$ of these coordinates. Equations of motion for $\bm \xi(t)$ can be derived along the same lines from the Landau-Lifshitz equation (\ref{eq:LLG}). They read \cite{Clarke:2008} 
\begin{equation}
\sum_j F_{ij} \dot{\xi}^j - \partial U/\partial \xi^i = 0,    
\label{eq:eom-collective-coords}
\end{equation}
where the potential-energy functional $U[\mmm(\rrr,\bm \xi)]$ is treated as a function of the collective coordinates $U(\bm \xi)$.

As in Thiele's equation, Eq.~(\ref{eq:eom-collective-coords}) expresses the balance of (generalized) forces for each coordinate $\xi^i$. The second term is a conservative force stemming from the dependence of the potential energy on coordinate $\xi^i$. The first term is the gyroscopic force, an analog of the Lorentz force. The proportionality constant between velocity $\dot{\xi}^j$ and the generalized force in the $\xi^i$ channel is the gyroscopic coefficient 
\begin{equation}
F_{ij}
= - 
\int dV \, \mathcal S
\mmm \cdot 
\left(
\frac{\partial \mmm}{\partial \xi^i}
\times 
\frac{\partial \mmm}{\partial \xi^j}
\right)
 = - F_{ji}.
\label{eq:gyroscopic-coefficient}
\end{equation}
The gyroscopic coefficients $F_{ij}$ form an antisymmetric second-rank tensor. 

We can see that Thiele's gyrovector (\ref{eq:gyrovector}) is a particular instance of the gyroscopic tensor: for a rigidly moving soliton, $\bm \xi = \{X, Y, Z\}$ and one obtains $G^i = - \frac{1}{2}\sum_{j,k} \epsilon_{ijk}  F_{ij}$ \cite{Thiele:1973a}.

For future reference, we express the gyroscopic tensor in terms of spin spherical angles, 
\begin{equation}
F_{ij} = \int dV \, \mathcal S 
\left(
\frac{\partial \cos{\theta}}{\partial \xi^i}
\frac{\partial \phi}{\partial \xi^j}
- 
\frac{\partial \phi}{\partial \xi^i}
\frac{\partial \cos{\theta}}{\partial \xi^j}
\right).
\label{eq:gyroscopic-coefficient-theta-phi}
\end{equation}

\section{Gyroscopic tensor as a magnetic field}
\label{sec:F-magnetic-field}

The analogy with the motion of a massless charged particle in a background magnetic, noted in Sec.~\ref{sec:translational-motion}, extends to Eq.~(\ref{eq:eom-collective-coords}) as well. Parameters $\bm \xi$ can be thought of as coordinates of a particle in an $N$-dimensional manifold with a background magnetic field represented by a vector gauge potential with components $A_i(\bm \xi)$. Generally in $N$ dimensions, the magnetic field is an antisymmetric second-rank tensor $F_{ij}$, whose components are given by the curl of the vector potential, 
\begin{equation}
F_{ij}  = - F_{ji}
= \frac{\partial A_j}{\partial \xi^i}   
- \frac{\partial A_i}{\partial \xi^j}.
\label{eq:F-to-A}
\end{equation}
Only in $N=3$ dimensions, an antisymmetric second-rank tensor can be expressed in terms of a vector, $F_{ij} = - \epsilon_{ijk} B^k$, to yield the familiar description of a magnetic field as a vector $\mathbf B$.  

An important constraint imposed on any magnetic field that comes from a gauge potential, as described by Eq.~(\ref{eq:F-to-A}), is the Bianchi identity, 
\begin{equation}
\frac{\partial F_{jk}}{\partial \xi^i}
+ \frac{\partial F_{ki}}{\partial \xi^j}
+ \frac{\partial F_{ij}}{\partial \xi^k}
= 0.
\label{eq:Bianchi}
\end{equation}
In a Euclidean space of 3 dimensions, the Bianchi identity yields one of Maxwell's equations, $\nabla \cdot \mathbf B = 0$, expressing the absence of sources and sinks of the magnetic field. 

It is straightforward to check that the gyroscopic tensor (\ref{eq:gyroscopic-coefficient}) satisfies the Bianchi identity (\ref{eq:Bianchi}). (It helps to remember that the spin field $\mmm$ has unit length, so any derivative of $\mmm$ is orthogonal to $\mmm$. For this reason, the three vectors $\partial \mmm/\partial \xi^i$, $\partial \mmm/\partial \xi^j$, and $\partial \mmm/\partial \xi^k$ are coplanar and their triple product vanishes.)

\section{Examples}
\label{sec:examples}

To illustrate these general considerations, we discuss a couple of examples. 

\subsection{Domain wall in a ferromagnetic wire}
\label{sec:examples-wire}

In a thin ferromagnetic wire, magnetostatic dipolar interactions alone induce easy-axis shape anisotropy, resulting in two ground states with the magnetization field parallel to the wire. For a magnetic wire stretched along the $z$-axis, the ground states are $\mmm(z) = \pm \mathbf e_z$. 

\begin{figure}
\centering
\includegraphics[width=0.7\columnwidth]{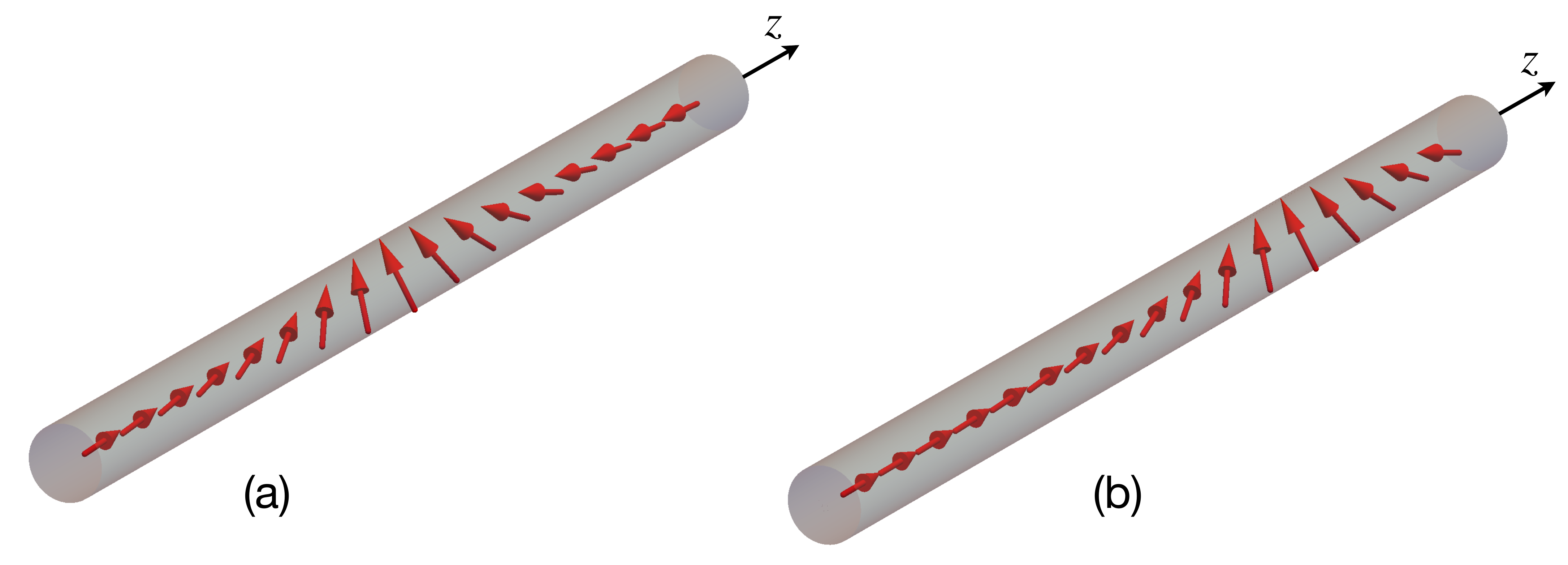}
\caption{Domain wall with the $\mathbb Z_2$ topological charge $\zeta = -1$ in a ferromagnetic wire. (a) A domain wall positioned in the middle of the wire, $Z = 0$, has angular momentum $P_\Phi = 0$. (b) A domain wall centered at $Z$ has $P_\Phi = - 2 \zeta \mathcal S Z$.}
\label{fig:domain-wall}
\end{figure}

In addition to these global energy minima, the energy functional has two classes of local energy minima describing two domains of uniform magnetization separated by a domain wall. The precise configurations depend on the details of the energy functional, which may include intrinisic material anisotropy aside from the shape-induced one. However, the general features are universal. Expressed in terms of the spherical angles $\theta$ and $\phi$, the magnetization field is
\begin{equation}
\cos{\theta(z)} = \zeta f(z-Z), 
\quad
\phi(z) = \Phi.
\label{eq:wire-domain-wall}
\end{equation}
Here $f(z)$ is an odd function interpolating between $f(-\infty) = -1$ and $f(+\infty) = +1$; $\zeta = \pm 1$ is a topological charge distinguishing between ``head-to-head'' and ``tail-to-tail'' domain walls, Fig.~\ref{fig:domain-wall}. 

The simplest model of this kind has the energy functional that includes the exchange energy and local anisotropy \cite{Thiele:1973b, Yan:2011}: 
\begin{equation}
U[\mmm(z)] = 
\int dz \,
\left[
A(\partial_z\mmm)^2
+ K (m_x^2 + m_y^2)
\right]/2.
\label{eq:wire-energy}    
\end{equation}
In this model, the profile of the longitudinal magnetization is 
\begin{equation}
f(z) = \tanh{(z/\lambda)}.    
\end{equation}
The width of a domain wall is $\lambda = \sqrt{A/K}$; its energy is $E = 2\sqrt{AK}$. 

A generic model of the easy-axis ferromagnet has a discrete symmetry of time reversal, 
\begin{equation}
\mmm(z) \mapsto -\mmm(z),    
\label{eq:symmetry-time-reversal}
\end{equation}
and two continuous symmetries: Translations 
\begin{equation}
\mmm(z) \mapsto \mmm(z-Z)   
\label{eq:symmetry-translation}
\end{equation}
and global spin rotations.
\begin{equation}
\theta(z) \mapsto \theta(z),
\quad
\phi(z) \mapsto \phi(z) + \Phi.
\label{eq:symmetry-spin-rotation}
\end{equation}
The time-reversal symmetry is spontaneously broken by the ground states; the other two symmetries are preserved by them.

The domain-wall solutions (\ref{eq:wire-domain-wall}) spontaneously break all the symmetries. The breaking of the continuous symmetries results in the emergence of zero modes, i.e., moves with zero energy cost. It is not a coincidence that the domain-wall coordinates $Z$ and $\Phi$ have the same names as the translation and rotation parameters in Eqs.~(\ref{eq:symmetry-translation}) and (\ref{eq:symmetry-spin-rotation}). 

Being local energy minima, the domain-wall solutions have a vanishing functional derivative $\delta U/\delta\mmm(z)$ and are therefore static in the absence of symmetry-perturbations. A weak perturbation will leave the domain-wall solutions largely intact and introduce  some energy landscape $U(Z,\Phi)$. This will create conservative forces and induce the motion of domain walls. Eq.~(\ref{eq:eom-collective-coords}) reads
\begin{equation}
F_{Z\Phi} \dot{\Phi} - \partial U/\partial Z = 0,
\quad
F_{\Phi Z} \dot{Z} - \partial U/\partial \Phi = 0.
\end{equation}
The nonvanishing components of the gyroscopic tensor are
\begin{equation}
F_{Z\Phi} = - F_{\Phi Z} = - 2 \zeta \mathcal S.
\label{eq:F-domain-wall}
\end{equation}
It is worth noting that the gyroscopic coefficients depend on the topological charge of the domain wall $\zeta$ but not on its precise shape $f(z)$. The only other quantity appearing in Eq.~(\ref{eq:F-domain-wall}) is the spin density $\mathcal S$ (spin per unit length).

A more realistic model, where the axial symmetry is broken by additional anisotropy favoring the $m_y$ spin component over $m_x$, was studied by Schryer and Walker \cite{Schryer:1974}. The width of the domain wall becomes dependent on the azimuthal angle $\Phi$. Still, the gyroscopic coefficients (\ref{eq:F-domain-wall}) remain unchanged. 

\subsection{Vortex in a thin magnetic film}
\label{sec:examples-film}

\begin{figure}
\centering
\includegraphics[width=0.8\columnwidth]{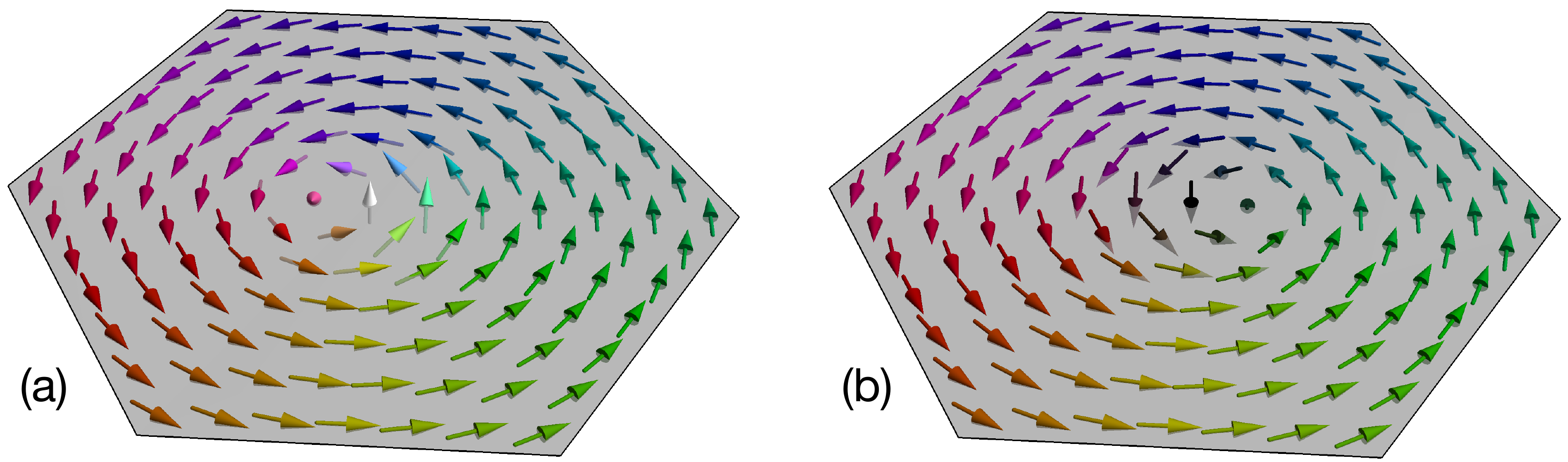}
\caption{Vortex in a thin ferromagnetic film. (a) $\chi = 1$, $p = 1$, and $Q=1/2$. (b) $\chi = 1$, $p=-1$, and $Q = -1/2$. Reprinted with permission from Ref.~\cite{Hellman:2017etal}.}
\label{fig:vortex}
\end{figure}

In a thin ferromagnetic film, magnetostatic dipolar interactions induce easy-plane shape anisotropy. With the film in the $xy$ plane, the uniform ground states have 
\begin{equation}
\cos{\theta(\rrr)} = 0, 
\quad
\phi(\rrr) = \Phi.
\end{equation}
Topological defects in this case are vortices and antivortices. A vortex centered at the origin has a shape
\begin{equation}
\cos{\theta(\rrr)} = p g(r),
\quad
\phi(\mathbf r) = \alpha + \chi \pi/2.
\quad
\label{eq:vortex-at-origin}
\end{equation}
Here $r$ and $\alpha$ are polar coordinates in the plane of the film,
\begin{equation}
x = r \cos{\alpha}, 
\quad 
y = r \sin{\alpha},
\end{equation}
$g(r)$ is a function interpolating between $g(0) = 1$ and $g(\infty) = 0$, $p = \pm 1$ is the polarity of the vortex and $\chi = \pm 1$ its chirality. See Fig.~\ref{fig:vortex}. Rigid translations of the vortex solution (\ref{eq:vortex-at-origin}) yield states with the same energy, 
\begin{equation}
\mmm(\rrr) \mapsto \mmm(\rrr-\RRR). 
\end{equation}

Translations are the only zero modes of a vortex. Taking $\RRR=(X,Y)$ as the collective coordinates of a vortex, we obtain their gyroscopic coefficients 
\begin{equation}
F_{XY} = - F_{YX} = - 4\pi Q \mathcal S,   
\label{eq:F-vortex}
\end{equation}
where 
\begin{equation}
Q = \frac{1}{4\pi} \int d^2r \, 
\mmm \cdot (\partial_x \mmm \times \partial_y \mmm)
= p/2
\end{equation}
is a topological charge known as the skyrmion number \cite{Tretiakov:2007}. It quantifies the degree of mapping $\rrr \mapsto \mmm$, counting the number of times the plane maps onto the sphere. The field $\mmm(\rrr)$ of a vortex covers the northern ($p=+1$) or southern ($p=-1$) hemisphere. Therefore, the skyrmion number of a vortex is half-integer, $Q=\pm 1/2$. 

Once again, the gyroscopic coefficients are insensitive to the precise shape $g(r)$ of the vortex and depend only on the spin density $\mathcal S$ (spin per unit area) and a topological charge. 

\subsection{Robust nature of the gyroscopic tensor}

The two examples considered in this section show that the gyroscopic coefficients are very robust characteristics of a soliton defined not by its shape but rather by its topology. This remarkable stability hints that perhaps the gyroscopic tensor is not just an auxiliary quantity used in calculating the gyroscopic force but is instead a more fundamental object underpinning much of the mechanics of magnetic solitons. 

\section{Conserved momenta of ferromagnetic solitons}
\label{sec:conserved-momenta}

Continuous symmetries are directly associated with conservation laws. The model of a magnetic wire in Sec.~\ref{sec:examples-wire} provides a salient example. Translational symmetry (\ref{eq:symmetry-translation}) gives rise to a conserved linear momentum $P_z$; the symmetry of global spin rotations (\ref{eq:symmetry-spin-rotation}) ensures the conservation of angular momentum $J_z$. 

The angular momentum is easy to visualize. As a one-dimensional object with no $x$ and $y$ coordinates, a wire has no orbital contribution to the angular momentum along the $z$ axis. Thus $J_z$ must come from spins alone. In a finite wire extending from $z=-L/2$ to $z=L/2$, with the domain wall centered in the middle of the wire, Fig.~\ref{fig:domain-wall}(a), equal numbers of spins point in the positive and negative $z$ directions so that the net angular momentum $J_z$ vanishes for this configuration. Moving the domain wall from the center of the wire by $Z$ elongates the $m_z=-\zeta$ domain by $Z$ and shortens the $m_z=+\zeta$ by the same amount, Fig.~\ref{fig:domain-wall}(b). The net angular momentum becomes $J_z = - 2\zeta \mathcal S Z$, where $\mathcal S$ is the spin density per unit length. Because this conserved momentum is conjugate to angular coordinate $\Phi$, we will refer to it as $P_\Phi$: 
\begin{equation}
P_\Phi = - 2\zeta \mathcal S Z. 
\label{eq:P-Phi-domain-wall}
\end{equation}

The linear momentum of a domain wall is not so easy to picture. Haldane derived the linear momentum of a one-dimensional ferromagnet from the energy-momentum tensor \cite{Haldane:1986}. This recipe yields the following result for the linear momentum of a domain wall \cite{Kosevich:1990, Galkina:2008, Yan:2013}: 
\begin{equation}
P_Z = 2\zeta \mathcal S \Phi. 
\label{eq:P-Z-domain-wall}
\end{equation}

It is remarkable that both the angular (\ref{eq:P-Phi-domain-wall}) and linear (\ref{eq:P-Z-domain-wall}) momenta are proportional to the gyroscopic coefficients of the domain wall (\ref{eq:F-domain-wall}). This is not a mere coincidence. It turns out that, quite generally, the conserved momentum for any collective coordinate $\xi^i$ whose translations are a symmetry can be expressed directly in terms of the gyroscopic tensor \cite{Tchernyshyov:2015}: 
\begin{equation}
P_i(2) - P_i(1) = - \int_1^2 \sum_j F_{ij}(\bm \xi) d\xi^j.    
\label{eq:P-through-F}
\end{equation}
Here the difference of conserved momenta at points 1 and 2 on the left is expressed in terms of a line integral in the space of collective coordinates. The results for the angular (\ref{eq:P-Phi-domain-wall}) and linear (\ref{eq:P-Z-domain-wall}) momenta are particular instances of the general result (\ref{eq:P-through-F}). 

Note that Eq.~(\ref{eq:P-through-F}) resembles the definition of potential energy in terms of a line integral of a force,
\begin{equation}
U(2) - U(1) = 
- \int_{1}^{2} \mathbf F(\mathbf r) \cdot d\rrr.
\end{equation}
Potential energy is only well defined for conservative forces, whose line integral depends on the endpoints but not the path between them. A necessary condition for that is the vanishing of the curl for the force, $\nabla \times \mathbf F = 0$. The counterpart of the zero-curl condition for the gyroscopic tensor is the Bianchi identity (\ref{eq:Bianchi}). If it holds then we are guaranteed that the line integral in Eq.~(\ref{eq:P-through-F}) is the same for any two paths (with the same endpoints) that can be continuously deformed into each other \cite{Tchernyshyov:2015}. 

The linear momentum of a vortex or a skyrmion in a thin film, first derived by Papanicolaou and Tomaras \cite{Papanicolaou:1991}, can also be obtained from the gyroscopic tensor (\ref{eq:F-vortex}) with the aid of Eq.~(\ref{eq:P-through-F}): 
\begin{equation}
P_X = 4\pi \mathcal S Q Y, 
\quad
P_Y = - 4\pi \mathcal S Q X.
\end{equation}
Note that, just like for a domain wall, the momenta are not independent of the coordinates. It is a general feature of precessional dynamics that the phase space of a ferromagnetic soliton coincides with its configuration space (\ref{eq:xi-collective-coordinates}). 

We end this section with two remarks. First, we have found that the gyroscopic tensor of a ferromagnetic soliton is directly, and very simply, related to the most fundamental physical characteristics of the soliton---its conserved momenta. This relation is expressed by Eq.~(\ref{eq:P-through-F}). Second, we have revealed that the obscure Bianchi identity (\ref{eq:Bianchi}) plays an essential role in soliton mechanics. It ensures that conserved momenta of a soliton, expressed in terms of a line integral (\ref{eq:P-through-F}), are well defined.

\section{Counting physical states}
\label{sec:counting-states}

Another fundamental aspect of soliton physics where the gyroscopic tensor plays a significant role is counting physical states. To see why this might be the case, consider the related topic of the quantum Hall effect, where electrons are confined to a thin layer and move in two spatial dimensions in a magnetic field normal to their plane. The strength of the field, expressed as the area density of magnetic flux quanta $\Phi_0 = 2\pi \hbar c/e$, sets the density of physical states accessible to an electron in a given Landau level \cite{Landau:1981}. In Sec.~\ref{sec:F-magnetic-field}, we compared the magnetic soliton to an electric charge and the gyroscopic tensor to a magnetic field. This analogy makes it plausible that a suitably quantified strength of the gyroscopic tensor may determine the density of physical states in the space of collective coordinates of a soliton. 

Once again, a domain wall in a magnetic wire (Sec.~\ref{sec:examples-wire}) provides an illuminating example. Its two soft modes, translations and global spin rotations, are parametrized by the collective coordinates $Z$ and $\Phi$. Its other modes are spin waves, whose frequency spectrum has a gap below some $\omega_\text{min}$. At energies below $\hbar \omega_\text{min}$, the quanta of spin waves (magnons) are frozen out, so that only the soft modes should be considered. In the toy model with local easy-axis anisotropy (\ref{eq:wire-energy}), $\omega_\text{min} = K /\mathcal S$ \cite{Thiele:1973b}. 

To get the count of the physical states associated with the soft modes $Z$ and $\Phi$, we use our previously obtained result (\ref{eq:P-Phi-domain-wall}), which states that, up to a multiplicative factor, spatial position $Z$ is the momentum for the angle $\Phi$. Therefore, the two do not commute and cannot be known simultaneously. We will use the eigenstates of angular momentum $P_\Phi = - 2\zeta \mathcal S Z$, which also happen to be eigenstates of position $Z$. Shifting the domain wall's position by $\Delta Z$ increases its angular momentum by $\Delta P_\Phi = - 2\zeta \mathcal S \Delta Z$. Quantum mechanics tell us that angular momentum can be incremented in steps of $\hbar$. Hence
\begin{equation}
\mathcal N 
= |\Delta P_\Phi|/\hbar
= 2 \mathcal S \Delta Z/\hbar
\label{eq:N-Z-Phi-quantum}
\end{equation}
is the number of states for a domain wall positioned between $Z$ and $Z+\Delta Z$. 

The same count can be obtained in a classical calculation. Since, up to a multiplicative factor, $Z$ and $\Phi$ are momenta of each other, the $(Z,\Phi)$ manifold can be thought of as a phase space, where the number of physical states is proportional to the area. The proportionality constant can be fixed with the aid of our quantum result (\ref{eq:N-Z-Phi-quantum}):
\begin{equation}
\mathcal N = \int \frac{d Z \, d \Phi} {2\pi \hbar} \sqrt{-F_{Z\Phi} F_{\Phi Z}},   
\end{equation}
with the gyroscopic coefficients from  (\ref{eq:F-domain-wall}).

More generally, one may judiciously choose $N$ collective coordinates (\ref{eq:xi-collective-coordinates}) to represent soft modes of a magnetic soliton, with the rest of the (hard) modes frozen out below some energy scale. The number of physical states in this $N$-dimensional configuration space can be expressed with the aid of the gyroscopic coefficients $F_{ij}$ of soft modes comprising an $N \times N$ antisymmetric matrix $F$: 
\begin{equation}
\mathcal N = 
\int \frac{d\xi^1 \ldots d\xi^N}{(2\pi \hbar)^{N/2}}
\sqrt{\det{F}}.
\label{eq:N-generic}
\end{equation}
The number of collective coordinates $N$ must be even; otherwise $\det{F} = 0$.

Eq.~(\ref{eq:N-generic}) is a new result as far as we know. It can be anticipated from the observation that the volume element $d\xi^1 \ldots d\xi^N \sqrt{\det{F}}$ is invariant under arbitrary non-degenerate transformations of coordinates. Here the gyroscopic tensor $F$ plays the same role as the metric tensor $g$ does in the spacetime volume element $dx^0 dx^1 dx^2 dx^3 \sqrt{|\det{g}|}$ in general relativity \cite{Landau:1980}. 

\section{Canonical quantization}

The gyroscopic tensor also comes in handy in the canonical quantization of soliton dynamics. We have already seen in Sec.~\ref{sec:conserved-momenta} that it relates collective coordinates to conserved momenta. Furthermore, it determines the structure of the phase space and thus underlies the canonical quantization procedure.

In technical terms, the gyroscopic tensor is the inverse of the Poisson bracket, see Ref. \cite{Tchernyshyov:2015} and Sec.~\ref{sec:history} below. For a domain wall of Sec.~\ref{sec:examples-wire}, with the collective coordinates restricted to position $Z$ and azimuthal angle $\Phi$ (i.e., no spin waves), the matrix of Poisson brackets is 
\begin{equation}
\left(
\begin{array}{cc}
    \{Z,Z\} & \{Z,\Phi\} \\
    \{\Phi,Z\} & \{\Phi,\Phi\}
\end{array}
\right)
= \left(
\begin{array}{cc}
    F_{ZZ} & F_{Z\Phi} \\
    F_{\Phi Z} & F_{\Phi\Phi}
\end{array}
\right)^{-1}
= \left(
\begin{array}{cc}
    0 & 1/2 \zeta \mathcal S  \\
    -1/2 \zeta \mathcal S & 0
\end{array}
\right),
\end{equation}
where we have used Eq.~(\ref{eq:F-domain-wall}). Canonical quantization boils down to promoting physical observables to operators and their Poisson brackets to commutators: 
\begin{equation}
[\hat{Z}, \hat{\Phi}] = i \hbar/2 \zeta \mathcal S. 
\end{equation}

If a domain wall is perturbed by a weak transverse magnetic field $\mathbf h = (h,0,0)$ the perturbation Hamiltonian is 
\begin{equation}
\hat{V} = - \pi \mathcal S \gamma h \lambda   \cos{\hat{\Phi}},    
\end{equation}
where $\gamma$ is the gyromagnetic ratio. With translational invariance preserved, soliton momentum $\hat{P}_Z = 2 \zeta \mathcal S \hat{\Phi}$ commutes with the Hamiltonian. For a ferromagnetic chain with a lattice spacing $a$ and spins of length $\hbar S$, the spin density is $\mathcal S = \hbar S/a$ and the energy dispersion of a domain wall is 
\begin{equation}
E(P_Z) = 2\sqrt{AK}
- \pi \hbar S \gamma h \frac{\lambda}{a} 
\cos{\frac{P_Z a}{2\hbar S}}.    
\label{eq:E-P-domain-wall}
\end{equation}

To put this result in a familiar context, consider the extreme quantum version of the easy-axis ferromagnet in one dimension, the Ising chain in a transverse magnetic field \cite{Pfeuty:1970, Sachdev:2011}. It has spins of length $S=1/2$ with the Hamiltonian
\begin{equation}
\hat{H} = - \sum_n 
[J \hat{s}_n^z \hat{s}_{n+1}^z 
+ \hbar \gamma h \hat{s}_n^x].
\end{equation}
In a weak field, $\hbar \gamma h \ll J$, a domain wall with momentum $P_Z$ has the energy 
\begin{equation}
E(P_Z) = 
\frac{J}{2} 
- \hbar \gamma h \cos{\frac{P_Z a}{\hbar}},
\end{equation}
which matches Eq.~(\ref{eq:E-P-domain-wall}) for $S=1/2$. 

Vortex motion (Sec.~\ref{sec:examples-film}) can be quantized along similar lines. From the gyroscopic coefficients for vortex position $X$ and $Y$ (\ref{eq:F-vortex}) we immediately obtain their Poisson bracket $\{X,Y\} = 1/4\pi Q \mathcal S$ and the commutator for the corresponding operators,
\begin{equation}
[\hat{X},\hat{Y}] 
= \frac{i \hbar }{4\pi Q \mathcal S}
= \frac{i a^2}{4\pi Q S}    
\end{equation} 
in a square lattice with a lattice period $a$ and spins of length $\hbar S$. A weak pinning potential from the atomic lattice creates a perturbation Hamiltonian \cite{Galkin:2007, Takashima:2016}
\begin{equation}
\hat{V} = 
g
\left(
\cos{\frac{2\pi \hat{X}}{a}}
+ \cos{\frac{2\pi \hat{Y}}{a}}
\right).   
\label{eq:perturbation-vortex}
\end{equation}
This Hamiltonian was first diagonalized in a different but related context: the energy bands of an electron in a periodic lattice in a uniform magnetic field \cite{Azbel:1964, Hofstadter:1976}. 

\section{Deep history}
\label{sec:history}

The first example of a gyroscopic tensor for a magnetic soliton was introduced in 1973 by Thiele \cite{Thiele:1973a} for translational motion. Clarke et al. \cite{Clarke:2008} generalized it to arbitrary collective coordinates in 2008. However, essentially the same concept, albeit in a different context, was invented in 1808 by Lagrange \cite{Lagrange:1877, Marle:2009}.  

Lagrange was interested in celestial mechanics. To a zeroth approximation, the motion of planets is described by the Kepler model, where a planet orbits the sun. A planet has 3 coordinates $x$, $y$, $z$ and 3 velocities $\dot{x}$, $\dot{y}$, $\dot{z}$. There are 6 integrals of motion: 3 components of the angular momentum and 3 components of the Runge-Lenz vector. (The seventh integral of motion, the energy, can be expressed in terms of these.) These 6 constants of motion fully determine the orbit of the planet around the sun. Astronomers use a more practical set of constants of motion, known as the orbital elements: the orientation of the orbital plane (2 angles), the orientation of the orbit's major semiaxis in the orbital plane (1 angle), the eccentricity of the elliptical orbit, the length of the major axis, and the initial position along the orbit.

Perturbations beyond the Kepler model, such as the gravitational tug of other planets, cause the orbital elements to change with time. A well-known example is the precession of Mercury's orbit. Lagrange derived the equations governing the dynamics of orbital elements $\xi$ \cite{Lagrange:1877, Marle:2009, Iglesias-Zemmour:2013}: 
\begin{equation}
\sum_j (\xi^i,\xi^j) \dot{\xi}^j - \partial U_1/\partial \xi^i = 0. 
\label{eq:Lagrange-equation-orbital-elements}
\end{equation}
Here $\xi^i(t)$, $i = 1, \ldots, 6$, are orbital elements; $U_1(\bm \xi,t)$ is the potential energy of the perturbation; $(\xi^i,\xi^j)$ is the Lagrange bracket 
\begin{equation}
(\xi^i,\xi^j) \equiv \sum_n 
\left(
\frac{\partial p_n}{\partial \xi^i}
\frac{\partial q^n}{\partial \xi^j}
- 
\frac{\partial p_n}{\partial \xi^j}
\frac{\partial q^n}{\partial \xi^i}
\right),
\label{eq:Lagrange-bracket}
\end{equation}
where $q^n$, $n = 1, 2, 3$, are (generalized) coordinates of the planet and $p_n = \partial L/\partial \dot{q}^n$ are their canonical momenta obtained from a Lagrangian $L(\mathbf q, \dot{\mathbf q})$. 

Lagrange's equation for orbital elements (\ref{eq:Lagrange-equation-orbital-elements}) looks very much like the equation of motion for collective coordinates of a magnetic soliton (\ref{eq:eom-collective-coords}). Our gyroscopic tensor is the counterpart of the Lagrange bracket. 

This is not a coincidence. In fact, we can see that the two quantities have the same mathematical structure. To that end, take the expression for the gyroscopic tensor in terms of the spherical angle (\ref{eq:gyroscopic-coefficient-theta-phi}) and pass from the continuum description to an atomic one by replacing the volume integral $\int dV \, \mathcal S \ldots$ with a sum over individual spins $\sum_n S \ldots$ Here $S$ is the length of a spin. We obtain 
\begin{equation}
F_{ij} = \sum_n 
\left(  
\frac{\partial S \cos{\theta_n}}{\partial \xi^i}
\frac{\partial \phi_n}{\partial \xi^j}
- \frac{\partial S \cos{\theta_n}}{\partial \xi^j}
\frac{\partial \phi_n}{\partial \xi^i}
\right),
\label{eq:gyroscopic-coefficient-discrete}
\end{equation}
For the $n$th atomic spin, we may take $\phi_n$ as its coordinate $q^n$. Then $S \cos{\theta_n} = S_{nz}$ is its canonical momentum $p_n$. Comparing Eqs.~(\ref{eq:Lagrange-bracket}) and (\ref{eq:gyroscopic-coefficient-discrete}) shows that the gyroscopic tensor is, indeed, the Lagrange bracket: $F_{ij} = (\xi^i,\xi^j)$. 

Lagrange's bracket is not widely known among physicists. For example, Goldstein writes that it is ``mainly of historical interest now'' \cite{Goldstein:2001}. Indeed, in theoretical physics it has been superseded by the Poisson bracket,
\begin{equation}
\{\xi^i,\xi^j\} \equiv \sum_{n} 
\left(
\frac{\partial \xi^i}{\partial q^n}
\frac{\partial \xi^j}{\partial p_n}
- 
\frac{\partial \xi^j}{\partial q^n}
\frac{\partial \xi^i}{\partial p_n}
\right).
\label{eq:Poisson-bracket}    
\end{equation}

The Lagrange and Poisson brackets are closely related. If the $\bm \xi$ manifold has the same number of dimensions as the phase space $(\mathbf q, \mathbf p)$ then the two brackets, viewed as second-rank tensors, are the inverses of each other: 
\begin{equation}
\sum_j \{\xi^i,\xi^j\}(\xi^j,\xi^k) = \delta_{ik}.    
\end{equation}
Thus the Poisson bracket can be used to solve Lagrange's equations of motion (\ref{eq:Lagrange-equation-orbital-elements}) for velocities $\dot{\xi}^i$. The result, 
\begin{equation}
\dot{\xi}^i 
= \sum_j \{\xi^i,\xi^j\} 
\partial U_1 / \partial \xi^j,
\end{equation}
can be put in a compact form familiar to us from Hamiltonian dynamics:
\begin{equation}
\dot{\xi}^i 
= \{\xi^i,U_1\} 
= \{\xi^i,H_0+U_1\},
\label{eq:Hamiltonian-dynamics}
\end{equation}
In the last transition, we used the invariance of $\xi^i$ under the dynamics governed by the Hamiltonian $H_0$ of the unperturbed problem, $\{\xi^i,H_0\} = 0$. For a historical account, see Refs.~\cite{Marle:2009, Iglesias-Zemmour:2013}. 

Lagrange's bracket has not entirely disappeared from mechanics. The modern description of Hamiltonian dynamics uses the language of symplectic geometry for the phase space $(\mathbf q, \mathbf p)$ \cite{Fecko:2006}. A fundamental geomertical object in phase space, the symplectic 2-form 
\begin{equation}
\omega = \sum_n dq^n \wedge dp_n, 
\label{eq:symplectic-form-p-q}
\end{equation}
can be expressed in $\bm \xi$ coordinates with the aid of the Lagrange bracket: 
\begin{equation}
\omega 
= \frac{1}{2} \sum_{i,j} (\xi^i,\xi^j) \, 
d\xi^i \wedge d\xi^j.
\label{eq:symplectic-form-xi}
\end{equation}
The closed nature of the symplectic form (\ref{eq:symplectic-form-p-q}), $d\omega = 0$, translates to the Bianchi identity (\ref{eq:Bianchi}) for the Lagrange bracket,
\begin{equation}
\frac{\partial (\xi^j, \xi^k)}{\partial \xi^i} 
+ \frac{\partial (\xi^k, \xi^i)}{\partial \xi^j} 
+ \frac{\partial (\xi^i, \xi^j)}{\partial \xi^k} = 0.
\label{eq:Bianchi-Lagrange-bracket}
\end{equation}
The counterpart of Eq.~(\ref{eq:Bianchi-Lagrange-bracket}) for the Poisson bracket is the Jacobi identity,
\begin{equation}
\{\{\xi^j,\xi^k\},\xi^i\}
+ \{\{\xi^k,\xi^i\},\xi^j\}
+ \{\{\xi^i,\xi^j\},\xi^k\}
= 0.
\end{equation}
Symplectic geometry is also a suitable natural language for the dynamics of solitons in a ferromagnet \cite{Di:2021}.

\section{Discussion}

In this paper, we aimed to show that the gyroscopic tensor, originally introduced as an auxiliary quantity in the computation of forces acting on a ferromagnetic soliton, turns out to be a rather fundamental object in the mechanics of magnetic solitons. A first hint of its significance was provided by a remarkable simplicity and stability of the gyroscopic coefficients: they typically depend on the spin density of the material and some topological charge but not on the exact shape or size of the soliton. 

The importance of the gyroscopic tensor came into focus with the realization that it directly determines conserved momenta of a soliton \cite{Tchernyshyov:2015}. The simple and general relation (\ref{eq:P-through-F}) obviates the need for complex \cite{Papanicolaou:1991}, and even ``treacherous'' \cite{Thiele:1974}, derivations in specific cases. 

Furthermore, the gyroscopic tensor plays a major role in statistical physics: it provides a direct quantitative measure for the semiclassical density of physical states, Eq.~(\ref{eq:N-generic}). To the best of our knowledge, this has not been previously pointed out. 

The gyrosopic tensor will also be useful in quantum mechanics of magnetic solitons. Its direct (or, rather, inverse) relation to the Poisson bracket \cite{Tchernyshyov:2015} shows that it quantifies canonical relations in phase space, the information required for canonical quantization. 

Given all of the above, how has such an important mathematical object escaped the attention of theoretical physicists for so long? Well, it hasn't. Its earliest counterpart, the Lagrange bracket, was invented in 1808 as an auxiliary quantity in calculations of planetary motion \cite{Lagrange:1877}. Although its full significance had not been realized for a while, its discovery led Poisson to introduce his more famous bracket in 1809. Viewed as second-rank tensors, the two objects are simply the inverses of each other. 

Although the Lagrange bracket is rarely mentioned in today's physics textbooks, it remains very much a cornerstone of Hamiltonian dynamics. Its modern incarnation, the symplectic 2-form (\ref{eq:symplectic-form-xi}), determines the geometry of phase space \cite{Fecko:2006}. Connections to statistical and quantum physics are therefore quite natural.

Turning our gaze from the glorious past to the uncertain future, we ask: where else could the gyroscopic tensor be useful? One of the directions we plan to explore is a unification of spin and charge in spintronics. The interaction of electric currents with magnetization is a topic of both practical and theoretical interest. The spin-transfer torque exerted by a spin-polarized polarized electric current on a magnetic soliton was introduced by Berger \cite{Berger:1978} and formalized by Bazaliy et al. \cite{Bazaliy:1998}. A reciprocal effect of a moving soliton on the electric current, a spin electromotive force, was described by Barnes and Maekawa \cite{Barnes:2007}. As we pointed out earlier \cite{Dasgupta:2018}, it is natural to treat the electric charge flowing through a ferromagnet on the same footing as the collective coordinates of a magnetic soliton. Then both the spin-transfer torque and the spin electromotive force are seen as instances of the gyroscopic force with equal and opposite gyroscopic coefficients. Rules of symplectic geometry may provide important clues for spintronics, much like Riemannian geometry did for general relativity. 

\section*{Acknowledgments}

O.T. is grateful to Chia-Ling Chien for introducing him to magnetic solitons and for constant support and encouragement; to Yaroslaw Bazaliy, Gia-Wei Chern, David Clarke, Sayak Dasgupta, Xingjian Di, Se Kwon Kim, Imam Makhfudz, Derek Reitz, Valery Slastikov, Oleg Tretiakov, Yaroslav Tserkovnyak, Xiaofan Wu, and Shu Zhang for collaborations on this topic; to Boris Ivanov, Stavros Komineas, Volodymyr Kravchuk, Jonathan Robbins, and Denis Sheka for illuminating discussions. This work was supported as part of the Institute for Quantum Matter, an Energy Frontier Research Center funded by the U.S. Department of Energy, Office of Science, Basic Energy Sciences under Award No. DE-SC0019331.

\bibliography{main}

\begin{thebibliography}{10}
\expandafter\ifx\csname url\endcsname\relax
  \def\url#1{\texttt{#1}}\fi
\expandafter\ifx\csname urlprefix\endcsname\relax\def\urlprefix{URL }\fi
\expandafter\ifx\csname href\endcsname\relax
  \def\href#1#2{#2} \def\path#1{#1}\fi

\bibitem{Thiele:1973a}
A.~A. Thiele, Steady-state motion of magnetic domains, Phys. Rev. Lett. 30
  (1973) 230--233.
\newblock \href {https://doi.org/10.1103/PhysRevLett.30.230}
  {\path{doi:10.1103/PhysRevLett.30.230}}.

\bibitem{Clarke:2008}
D.~J. Clarke, O.~A. Tretiakov, G.-W. Chern, Y.~B. Bazaliy, O.~Tchernyshyov,
  Dynamics of a vortex domain wall in a magnetic nanostrip: Application of the
  collective-coordinate approach, Phys. Rev. B 78 (2008) 134412.
\newblock \href {https://doi.org/10.1103/PhysRevB.78.134412}
  {\path{doi:10.1103/PhysRevB.78.134412}}.

\bibitem{Lagrange:1877}
J.-L. Lagrange,
  \href{http://sites.mathdoc.fr/cgi-bin/oetoc?id=OE_LAGRANGE__6}{M{\'e}moire
  sur la th{\'e}orie g{\'e}n{\'e}rale de la variation des constantes
  arbitraires dans tous les probl{\'e}mes de la m{\'e}canique}, in: {\OE}uvres
  de Lagrange, Vol.~VI, Gauthier-Villars, Paris, 1877, pp. 771--805.
\newline\urlprefix\url{http://sites.mathdoc.fr/cgi-bin/oetoc?id=OE_LAGRANGE__6}

\bibitem{Brown:1978}
W.~F. Brown, Jr., Domains, micromagnetics, and beyond: Reminiscences and
  assessments, J. Appl. Phys. 49 (1978) 1937--1942.
\newblock \href {https://doi.org/10.1063/1.324811}
  {\path{doi:10.1063/1.324811}}.

\bibitem{Landau:1935}
L.~Landau, E.~Lifshits, On the theory of the dispersion of magnetic
  permeability in ferromagnetic bodies, Phys. Z. Sowjet. 8 (1935) 153,
  reprinted in \emph{Collected Papers of L. D. Landau}, Pergamon, New York, pp.
  101-114 (1965).
\newblock \href {https://doi.org/10.1016/B978-0-08-010586-4.50023-7}
  {\path{doi:10.1016/B978-0-08-010586-4.50023-7}}.

\bibitem{Thiele:1973b}
A.~A. Thiele, Excitation spectrum of magnetic domain walls, Phys. Rev. B 7
  (1973) 391--397.
\newblock \href {https://doi.org/10.1103/PhysRevB.7.391}
  {\path{doi:10.1103/PhysRevB.7.391}}.

\bibitem{Yan:2011}
P.~Yan, X.~S. Wang, X.~R. Wang, All-magnonic spin-transfer torque and domain
  wall propagation, Phys. Rev. Lett. 107 (2011) 177207.
\newblock \href {https://doi.org/10.1103/PhysRevLett.107.177207}
  {\path{doi:10.1103/PhysRevLett.107.177207}}.

\bibitem{Schryer:1974}
N.~L. Schryer, L.~R. Walker, The motion of $180^\circ$ domain walls in uniform
  dc magnetic fields, J. Appl. Phys. 45 (1974) 5406--5421.
\newblock \href {https://doi.org/10.1063/1.1663252}
  {\path{doi:10.1063/1.1663252}}.

\bibitem{Hellman:2017etal}
F.~Hellman, A.~Hoffmann, Y.~Tserkovnyak, G.~S.~D. Beach, E.~E. Fullerton,
  C.~Leighton, A.~H. MacDonald, D.~C. Ralph, D.~A. Arena, H.~A. D\"urr, et~al.,
  Interface-induced phenomena in magnetism, Rev. Mod. Phys. 89 (2017) 025006.
\newblock \href {https://doi.org/10.1103/RevModPhys.89.025006}
  {\path{doi:10.1103/RevModPhys.89.025006}}.

\bibitem{Tretiakov:2007}
O.~A. Tretiakov, O.~Tchernyshyov, Vortices in thin ferromagnetic films and the
  skyrmion number, Phys. Rev. B 75 (2007) 012408.
\newblock \href {https://doi.org/10.1103/PhysRevB.75.012408}
  {\path{doi:10.1103/PhysRevB.75.012408}}.

\bibitem{Haldane:1986}
F.~D.~M. Haldane, Geometrical interpretation of momentum and crystal momentum
  of classical and quantum ferromagnetic {Heisenberg} chains, Phys. Rev. Lett.
  57 (1986) 1488--1491.
\newblock \href {https://doi.org/10.1103/PhysRevLett.57.1488}
  {\path{doi:10.1103/PhysRevLett.57.1488}}.

\bibitem{Kosevich:1990}
A.~M. Kosevich, B.~A. Ivanov, A.~S. Kovalev, Magnetic solitons, Phys. Rep.
  194~(3-4) (1990) 117--238.
\newblock \href {https://doi.org/10.1016/0370-1573(90)90130-t}
  {\path{doi:10.1016/0370-1573(90)90130-t}}.

\bibitem{Galkina:2008}
E.~G. Galkina, B.~A. Ivanov, S.~Savel'ev, F.~Nori, Chirality tunneling and
  quantum dynamics for domain walls in mesoscopic ferromagnets, Phys. Rev. B 77
  (2008) 134425.
\newblock \href {https://doi.org/10.1103/PhysRevB.77.134425}
  {\path{doi:10.1103/PhysRevB.77.134425}}.

\bibitem{Yan:2013}
P.~Yan, A.~Kamra, Y.~Cao, G.~E.~W. Bauer, Angular and linear momentum of
  excited ferromagnets, Phys. Rev. B 88 (2013) 144413.
\newblock \href {https://doi.org/10.1103/PhysRevB.88.144413}
  {\path{doi:10.1103/PhysRevB.88.144413}}.

\bibitem{Tchernyshyov:2015}
O.~Tchernyshyov, Conserved momenta of a ferromagnetic soliton, Ann. Phys. (NY)
  363 (2015) 98--113.
\newblock \href {https://doi.org/10.1016/j.aop.2015.09.004}
  {\path{doi:10.1016/j.aop.2015.09.004}}.

\bibitem{Papanicolaou:1991}
N.~Papanicolaou, T.~Tomaras, Dynamics of magnetic vortices, Nucl. Phys. B 360
  (1991) 425--462.
\newblock \href {https://doi.org/10.1016/0550-3213(91)90410-y}
  {\path{doi:10.1016/0550-3213(91)90410-y}}.

\bibitem{Landau:1981}
L.~D. Landau, E.~M. Lifshitz, Quantum Mechanics: Non-Relativistic Theory, 4th
  Edition, Vol.~3 of Course of Theoretical Physics, Butterworth-Heinemann,
  Boston, 1981.

\bibitem{Landau:1980}
L.~D. Landau, E.~M. Lifshitz, The Classical Theory of Fields, 4th Edition,
  Vol.~2 of Course of Theoretical Physics, Butterworth-Heinemann, Boston, 1980.

\bibitem{Pfeuty:1970}
P.~Pfeuty, {The one-dimensional Ising model with a transverse field}, Ann.
  Phys. (NY) 57 (1970) 79--90.
\newblock \href {https://doi.org/10.1016/0003-4916(70)90270-8}
  {\path{doi:10.1016/0003-4916(70)90270-8}}.

\bibitem{Sachdev:2011}
S.~Sachdev, Quantim Phase Transitions, 2nd Edition, Cambridge University Press,
  Cambridge, 2011, {Chapter 5}.
\newblock \href {https://doi.org/10.1017/CBO9780511973765}
  {\path{doi:10.1017/CBO9780511973765}}.

\bibitem{Galkin:2007}
A.~Y. Galkin, B.~A. Ivanov, Semiclassical dynamics of vortices in {2D}
  easy-plane ferromagnets, J. Exp. Theor. Phys. 104 (2007) 775--791.
\newblock \href {https://doi.org/10.1134/s1063776107050123}
  {\path{doi:10.1134/s1063776107050123}}.

\bibitem{Takashima:2016}
R.~Takashima, H.~Ishizuka, L.~Balents, Quantum skyrmions in two-dimensional
  chiral magnets, Phys. Rev. B 94 (2016) 134415.
\newblock \href {https://doi.org/10.1103/PhysRevB.94.134415}
  {\path{doi:10.1103/PhysRevB.94.134415}}.

\bibitem{Azbel:1964}
{M. Ya. Azbel'},
  \href{http://jetp.ras.ru/cgi-bin/e/index/e/19/3/p634?a=list}{Energy spectrum
  of a conduction electron in a magnetic field}, Sov. Phys. JETP 19 (1964)
  634--645.
\newline\urlprefix\url{http://jetp.ras.ru/cgi-bin/e/index/e/19/3/p634?a=list}

\bibitem{Hofstadter:1976}
D.~R. Hofstadter, Energy levels and wave functions of {Bloch} electrons in
  rational and irrational magnetic fields, Phys. Rev. B 14 (1976) 2239--2249.
\newblock \href {https://doi.org/10.1103/PhysRevB.14.2239}
  {\path{doi:10.1103/PhysRevB.14.2239}}.

\bibitem{Marle:2009}
C.-M. Marle, The inception of symplectic geometry: the works of {Lagrange} and
  {Poisson} during the years 1808-1810, Lett. Math. Phys. 90 (2009) 3--21.
\newblock \href {https://doi.org/10.1007/s11005-009-0347-y}
  {\path{doi:10.1007/s11005-009-0347-y}}.

\bibitem{Iglesias-Zemmour:2013}
P.~Iglesias-Zemmour, {Lagrange et Poisson, sur la variation des constantes},
  in: Y.~Kosmann-Schwarzbach (Ed.), Sim{\'e}on-Denis Poisson, Les
  math{\'e}matiques au service de la science, Les {\'E}ditions de l'{\'E}cole
  polytechnique, {\'E}cole Polytechnique, 2013, pp. 281--289, eprint
  \href{https://hal.archives-ouvertes.fr/hal-01288532}{hal-01288532}.

\bibitem{Goldstein:2001}
H.~Goldstein, C.~P. Poole, Jr., J.~L. Safko, Classical Mechanics, 3rd Edition,
  Pearson, New York, 2001, {Chapter} 9.5.

\bibitem{Fecko:2006}
M.~Fecko, Differential Geometry and Lie Groups for Physicists, Cambridge
  University Press, 2006.
\newblock \href {https://doi.org/10.1017/CBO9780511755590}
  {\path{doi:10.1017/CBO9780511755590}}.

\bibitem{Di:2021}
X.~Di, O.~Tchernyshyov, Conserved momenta of ferromagnetic solitons through the
  prism of differential geometry, SciPost Phys. 11 (2021) 108.
\newblock \href {https://doi.org/10.21468/scipostphys.11.6.108}
  {\path{doi:10.21468/scipostphys.11.6.108}}.

\bibitem{Thiele:1974}
A.~A. Thiele, On the momentum of ferromagnetic domains, J. Appl. Phys. 47
  (1976) 2759--2760.
\newblock \href {https://doi.org/10.1063/1.323005}
  {\path{doi:10.1063/1.323005}}.

\bibitem{Berger:1978}
L.~Berger, Low-field magnetoresistance and domain drag in ferromagnets, J.
  Appl. Phys. 49 (1978) 2156--2161.
\newblock \href {https://doi.org/10.1063/1.324716}
  {\path{doi:10.1063/1.324716}}.

\bibitem{Bazaliy:1998}
Y.~B. Bazaliy, B.~A. Jones, S.-C. Zhang, Modification of the {Landau-Lifshitz}
  equation in the presence of a spin-polarized current in colossal- and
  giant-magnetoresistive materials, Phys. Rev. B 57 (1998) R3213--R3216.
\newblock \href {https://doi.org/10.1103/PhysRevB.57.R3213}
  {\path{doi:10.1103/PhysRevB.57.R3213}}.

\bibitem{Barnes:2007}
S.~E. Barnes, S.~Maekawa, Generalization of {Faraday's} law to include
  nonconservative spin forces, Phys. Rev. Lett. 98 (2007) 246601.
\newblock \href {https://doi.org/10.1103/PhysRevLett.98.246601}
  {\path{doi:10.1103/PhysRevLett.98.246601}}.

\bibitem{Dasgupta:2018}
S.~Dasgupta, O.~Tchernyshyov, Energy-momentum tensor of a ferromagnet, Phys.
  Rev. B 98 (2018) 224401.
\newblock \href {https://doi.org/10.1103/PhysRevB.98.224401}
  {\path{doi:10.1103/PhysRevB.98.224401}}.

\end{thebibliography}

\end{document}